\pdfoutput=1  
\documentclass{article}
\usepackage[preprint]{neurips_2026}
\usepackage[utf8]{inputenc}
\usepackage[T1]{fontenc}
\usepackage[colorlinks=true,linkcolor=black,citecolor={blue!55!black},urlcolor={blue!55!black}]{hyperref}
\usepackage{url}
\usepackage{booktabs}
\usepackage{amsmath,amssymb}
\usepackage{graphicx}
\usepackage{xcolor}
\usepackage{natbib}
\usepackage{caption}
\usepackage{subcaption}
\usepackage{titlesec}

\graphicspath{{figures/}}

\titlespacing*{\section}{0pt}{5pt plus 1pt minus 1pt}{2pt}
\titlespacing*{\subsection}{0pt}{4pt plus 1pt minus 1pt}{1pt}
\newcommand{\dNArt}{250}
\newcommand{\dNVid}{1000}
\newcommand{\dNEnc}{4}
\newcommand{\dKseeds}{4}
\newcommand{\dNCalib}{50}
\newcommand{\dNVal}{50}
\newcommand{\dNTest}{150}
\newcommand{\dTau}{0.76}
\newcommand{\dLambda}{0.50}
\newcommand{\dBandwidth}{1.07}
\newcommand{\dPCA}{50}
\newcommand{\dRedMeanR}{0.56}
\newcommand{\dRedMaxR}{1.00}
\newcommand{\dRedPairs}{10}
\newcommand{\dRedTotal}{21}
\newcommand{\dTopoBalAcc}{0.98}
\newcommand{\dTopoMacroF}{0.98}
\newcommand{\dTopoChance}{0.25}
\newcommand{\dOutProfile}{1.00}
\newcommand{\dOutDcu}{0.35}
\newcommand{\dMMProfile}{0.49}
\newcommand{\dMMDcu}{0.64}
\newcommand{\dAnisDir}{-0.14}
\newcommand{\dAnisIso}{0.41}

\newcommand{\dSemDcu}{0.50}
\newcommand{\dSemRtwoBase}{0.243}
\newcommand{\dSemRtwoFull}{0.240}
\newcommand{\dSemDelta}{-0.004}
\newcommand{\dSemClsPair}{0.79}
\newcommand{\dGapPoint}{0.054}
\newcommand{\dGapLo}{0.050}
\newcommand{\dGapHi}{0.057}
\newcommand{\dKernMean}{1.00}
\newcommand{\dKernBest}{0.95}
\newcommand{\dVmfKern}{0.47}

\newcommand{\dQcovDiv}{97}

\newcommand{\dQmissDiv}{56}
\newcommand{\dMsixDelta}{-0.011}
\newcommand{\dMsixLo}{-0.215}
\newcommand{\dMsixHi}{0.045}
\newcommand{\dRelThreePair}{1.00}
\newcommand{\dRelTwoPair}{1.00}
\newcommand{\dRelThreeDcu}{1.00}

\newcommand{\dRelThreeKern}{1.00}

\newcommand{\dCBPair}{0.80}
\newcommand{\dCBDcu}{0.80}
\newcommand{\dCBKern}{0.70}
\newcommand{\dCBMode}{0.35}
\newcommand{\dTempWithin}{0.016}
\newcommand{\dTempAcross}{0.289}
\newcommand{\dTempTraj}{0.144}
\newcommand{\dTempDom}{trajectory}

\newcommand{\dSelSemNBeat}{7}
\newcommand{\dSelRefProfGain}{0.088}
\newcommand{\dSelRefNBeat}{3}

\title{Toward Uncertainty Quantification in Modern Art}

\newcommand{\blfootnote}[1]{%
  \begingroup
  \renewcommand{\thefootnote}{}\footnote{#1}%
  \addtocounter{footnote}{-1}%
  \endgroup
}
\author{\neuripsauthor{%
  Tirtho Roy\textsuperscript{*,\textdagger,1}\quad
  Ushashi Bhattacharjee\textsuperscript{*,1}\quad
  Showrav Kumar Saha\textsuperscript{2}\quad
  Sayantan Chakraborty\textsuperscript{3}\quad
  Koushik Howlader\textsuperscript{1}\quad
  Tanusree Bhattacharjee\textsuperscript{1}\\[0.3em]
  \normalsize
  \textsuperscript{1}Iowa State University\quad
  \textsuperscript{2}Independent Researcher\quad
  \textsuperscript{3}University of Dhaka%
}}

\begin{document}
\maketitle
\blfootnote{\textsuperscript{*}Equal contribution.\quad\textsuperscript{\textdagger}Corresponding author: \texttt{tirtho@iastate.edu}.}

\begin{abstract}
Asked to animate the same modern artwork under different random seeds, a text to video model returns
visibly different films, one reading per seed. Because modern art is ambiguous by intent, this
disagreement is signal, not noise. Yet prevailing uncertainty quantification (UQ) collapses a set of
generations to a dispersion scalar that says how much the seeds differ but not how: it cannot tell a
compact interpretation from a dominant reading plus an outlier, two competing modes, or diffuse
instability, nor whether the set still contains a rendering faithful to the original. We present the
\textbf{first study of the structure of generative uncertainty for modern art animation}, and a
reusable \textbf{protocol for identifying source blind multiseed uncertainty}: a suite of seven
source blind and six reference aware estimators; a \emph{distributional profile} (robust spread,
outlier influence, explicit topology, multimodality, anisotropy, leave one seed influence, reference
coverage); a distribution model ablation (vMF, Kent, ACG, Student $t$, kernel, mixture); eight
identification questions; and an artwork level statistical protocol. We build the first corpus:
\dNArt{} modern artwork captions rendered by Wan2.1 14B under four seeds ($\dNVid{}$ videos) across
\dNEnc{} encoders, artworks withheld from generation. As a diagnostic the protocol succeeds: it
classifies seed set topology at balanced accuracy \dTopoBalAcc{} (chance \dTopoChance{}), isolates
the outlier configuration at AUROC \dOutProfile{} where a scalar reaches only \dOutDcu{}, and splits
high uncertainty artworks into reference covering ($n{=}\dQcovDiv{}$) and reference missing
($n{=}\dQmissDiv{}$) diversity, reliably from three seeds and across encoders. Two negative results
mark its limits: added structure does not beat pairwise dispersion at predicting semantic
disagreement (cross validated $R^2$ \dSemRtwoFull{} vs \dSemRtwoBase{}), and source blind uncertainty
tracks interpretation, not reference fidelity. We release the protocol, corpus, and harness.
\end{abstract}

\section{Introduction}
Text to video models can now set a still artwork in motion from a short description, and modern art
makes an unusually exacting subject: ambiguous by intent, a single piece licenses many defensible
readings. Render the same modern artwork under several seeds and the outputs diverge in earnest, each
seed committing to a distinct interpretation, and that divergence is precisely what a creative system
should measure rather than suppress. How a generator is \emph{uncertain} when it renders art has
nonetheless gone unexamined. The uncertainty estimators in common use (directional concentration,
pairwise dispersion, embedding variance, and semantic
entropy~\citep{kuhn2023semantic,farquhar2024semantic,chattopadhyay2026dcu}) each return a single
number. A scalar can state \emph{how much} the seeds disagree; it cannot state \emph{how} they
disagree. Four seeds that share an average pairwise distance may nonetheless form one compact
cluster, three concordant renderings beside a lone outlier, two coherent interpretations, or four
mutually distinct readings (Figure~\ref{fig:scalar}). For a system curating art these demand
different responses (accept, discard a seed, present alternatives, or abstain), yet a dispersion
score assigns them one value.

This paper is, to our knowledge, the \textbf{first systematic study of the structure of generative
uncertainty for modern art animation}. The operative question is not ``how uncertain is the
generator'' but ``\emph{what is the structure of that uncertainty, and can we recover it reliably
from a handful of seeds}.'' We answer it with a \textbf{protocol} for the source blind setting, in
which the human artwork behind a caption is withheld from generation and consulted only for
evaluation, so that no deployable score ever inspects the original. The protocol is deliberately
self contained and reusable: a baseline suite, a distributional profile that names five kinds of
structure, a distribution model ablation that gauges how much parametric machinery four seeds can
support, eight identification questions (RQ1 to RQ8), and an artwork level statistical protocol with
strict calibration hygiene.

To render the question answerable we construct the first corpus designed for it: \dNArt{} modern
artwork captions rendered by Wan2.1 14B~\citep{wan2025} under four seeds and embedded by \dNEnc{}
encoders, each paired with a source artwork held out of generation. Running the protocol end to end
on this corpus, we establish that:
\begin{itemize}\itemsep2pt
\item the source blind scalar estimators are strongly redundant (\dRedPairs{} of \dRedTotal{} pairs
redundant, maximum $|r|=\dRedMaxR{}$), so a single dispersion axis accounts for most of them (RQ1);
\item the distributional profile recovers structure a scalar cannot, classifying controlled topology
at balanced accuracy \dTopoBalAcc{} and flagging the outlier configuration at AUROC \dOutProfile{}
against \dOutDcu{} for DCU (RQ2);
\item uncertainty resolves cleanly into reference covering and reference missing diversity (RQ4, RQ5),
while the simple components remain reliable at four seeds (RQ7) and stable across encoders (RQ8);
\item two negative results add precision: the profile does not outperform pairwise dispersion on
semantic disagreement (RQ6), and source blind uncertainty does not gauge reference fidelity.
\end{itemize}
The lesson for modern art generation is that structure identification, not scalar magnitude, is where
source blind uncertainty carries information, and that this first protocol makes both findings
measurable and portable to other generators and collections.

\section{Related Work}
\textbf{Uncertainty in generative models.} Semantic entropy groups multiple generations into meaning
equivalence classes and takes the entropy of the resulting partition~\citep{kuhn2023semantic,
farquhar2024semantic}, while directional concentration uncertainty (DCU) fits a von Mises Fisher
distribution to normalized output embeddings and reports the inverse
concentration~\citep{chattopadhyay2026dcu}. Both compress the sample to a single scalar. We retain
them as baselines and ask, through objective tests, precisely what that scalar does and does not
capture.

\textbf{Text to video and its evaluation.} Wan2.1~\citep{wan2025} is a strong open text to video
model; corpus scale understanding~\citep{internvid2023} and benchmark suites~\citep{huang2024vbench}
appraise fidelity and text alignment, and world model research examines single
trajectories~\citep{mei2025worldmodels}. None characterizes the \emph{structure} of seed level
disagreement, which is our object of study.

\textbf{Creative agency and art.} Work on human and AI cocreation and creative
agency~\citep{doshi2025symbiosis,rafner2025agency,zhang2025agencylens,issak2025mosaaic,
xie2024contribution} argues for treating multiple generations as interpretations rather than errors,
and painting and animation systems~\citep{liu2025everypainting,hu2025animatepainter,
mahapatra2023cinemagraph} translate art into motion. None, however, asks how a generator is
\emph{uncertain} when it renders a modern artwork, that is, what the structure of its seed level
interpretation is. That measurement question, posed for modern art in particular, is the one we take
up first.

\section{The Protocol}
\label{sec:protocol}
\begin{figure}[t]
\centering
\includegraphics[width=0.6\textwidth]{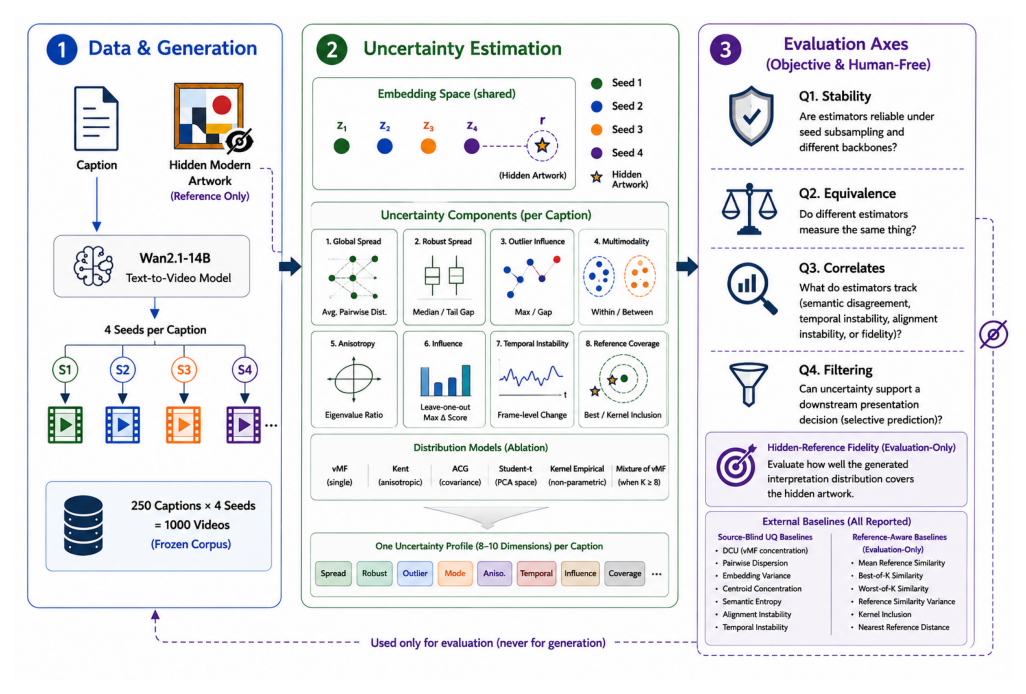}
\caption{The source blind protocol. The caption is derived from the artwork; the generator sees only
the caption and produces $K$ seeded videos whose embeddings form the interpretation distribution.
The artwork bypasses generation and enters only the evaluation branch.}
\label{fig:framework}
\end{figure}

\textbf{Setup and notation.} For artwork $i$ we observe $\mathcal D_i=(I_i,c_i,V_{i1},\dots,V_{iK})$:
a hidden reference image $I_i$, a caption $c_i$ derived from it, and $K$ videos $V_{ik}$ generated
from $c_i$ under distinct seeds. The artwork never enters generation. Each video is encoded frame by
frame and mean pooled to a unit clip vector $x_{ik}$, and the reference and caption are encoded by
the same tower to $\tilde r_i,\tilde q_i$. The $K$ clips constitute an empirical \emph{interpretation
distribution} $\widehat P_i^{\mathrm{gen}}$, which the protocol sets out to characterize
(Figure~\ref{fig:framework}).

\textbf{Baseline suite (source blind and reference aware).} Seven source blind estimators read the
videos alone: DCU (inverse vMF concentration), pairwise dispersion, embedding variance, centroid
concentration, meaning equivalence semantic entropy, caption alignment instability, and within video
temporal instability. Six reference aware measures consult the hidden artwork and are labeled
\emph{evaluation only}: the mean, best of $K$, worst of $K$, and variance of reference similarity,
kernel reference inclusion, and nearest seed distance. The boundary between the two families is never
crossed by a deployable score: only the reference aware measures ever read the withheld artwork, and
they serve evaluation exclusively.

\textbf{Distributional profile.} The core of the protocol decomposes $\widehat P_i^{\mathrm{gen}}$
into named components: global spread (pairwise dispersion); \emph{robust spread} (the median pairwise
distance and a tail gap $=\text{mean}-\text{median}$); \emph{outlier influence} (each seed's mean
distance to the rest, its maximum, and the gap between the two largest); an \emph{explicit topology}
obtained by selecting the seed partition $\hat{\mathcal C}_i=\arg\min_{\mathcal C}\,
S^{\mathrm{within}}_i(\mathcal C)+\lambda|\mathcal C|$ over the candidate shapes
$\{(4),(3{+}1),(2{+}2),(2{+}1{+}1),(1{+}1{+}1{+}1)\}$; \emph{multimodality} (the between over within
scatter ratio of the selected partition); \emph{anisotropy} (the leading eigenvalue share and
effective dimensionality in a shared PCA space); \emph{leave one seed influence}; and \emph{reference
coverage} (best of $K$ and a kernel inclusion score). The penalty $\lambda$, the kernel bandwidth
$h$, and the PCA basis are fit on the calibration split alone.

\textbf{Distribution model ablation.} To gauge how much parametric structure four seeds can support,
the protocol fits six models to each seed set, namely a single vMF, a Kent distribution with
shrinkage, an angular central Gaussian, a Student $t$ in PCA space, an empirical spherical kernel,
and a vMF mixture (used only when $K\!\ge\!8$), and scores each by held out (leave one seed) log
likelihood, reference inclusion likelihood, and stability under subsampling.

\textbf{Statistical protocol.} Every estimate resamples at the \emph{artwork} level, moving all seeds
together, across $10^4$ bootstrap replicates; correlations carry permutation $p$ values; method
comparisons use a paired artwork level bootstrap; and a Benjamini Hochberg correction is applied
within each experiment family. The data are split by artwork into \dNCalib{} calibration, \dNVal{}
validation, and \dNTest{} test, stratified by style. Calibration yields $\tau{=}\dTau{}$ (the
semantic threshold), $\lambda{=}\dLambda{}$ (the topology penalty), $h{=}\dBandwidth{}$ (the kernel
bandwidth), and a \dPCA{} dimensional PCA space.

\begin{figure}[t]
\centering
\includegraphics[width=\textwidth]{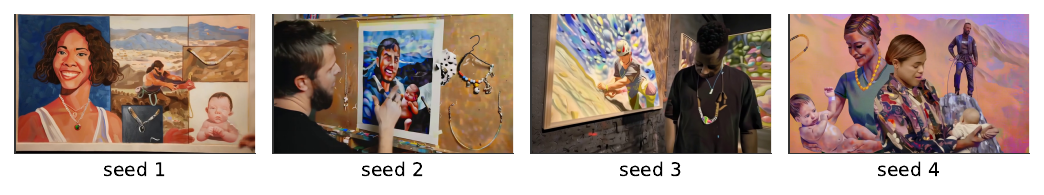}
\caption{One caption rendered under four seeds (one representative frame per seed). Each seed advances
a distinct reading of the same modern artwork, so the four clips constitute four samples of the
interpretation distribution the protocol characterizes.}
\label{fig:filmstrip}
\end{figure}

\section{A Modern Art Rendering Corpus}
We build the first corpus for structural uncertainty in modern art animation: \dNArt{} modern artwork
captions spanning four style strata (postimpressionism, pop art, surrealism, and a modern generic
set), each rendered by Wan2.1 14B~\citep{wan2025} at $832\times480$, 24 frames, under
$K{=}\dKseeds{}$ seeds, for a frozen corpus of \dNVid{} videos (Figure~\ref{fig:filmstrip}). No new generation is performed; the
protocol runs entirely off cached embeddings. Videos, references, and captions are embedded by
\dNEnc{} encoders, OpenCLIP ViT B/32 (primary) together with ViT L/14, SigLIP, and DINOv2 (vision
only), so that every ranking can be tested for encoder dependence. Structured cross seed semantic
disagreement over subject, setting, action, style, mood, color, and composition is elicited from a
source blind vision language annotator applied to seed mid frames, supplying an observable target
that never sees the reference.

\section{Applying the Protocol}
\subsection{RQ1: Are the scalar estimators redundant?}
Across the source blind estimators the mean absolute pairwise correlation is \dRedMeanR{}, and
\dRedPairs{} of \dRedTotal{} pairs are redundant ($|r|>0.8$), reaching a maximum of \dRedMaxR{}. DCU,
pairwise dispersion, embedding variance, and centroid concentration behave as near monotone
transforms of one another. \emph{Conclusion:} the source blind scalars occupy essentially one
dispersion axis, and a further dispersion scalar contributes little.

\subsection{RQ2: Does the profile recover structure a scalar cannot?}
\label{sec:recovery}
\begin{figure}[t]
\centering
\begin{subfigure}[t]{0.4\textwidth}
\includegraphics[width=\textwidth]{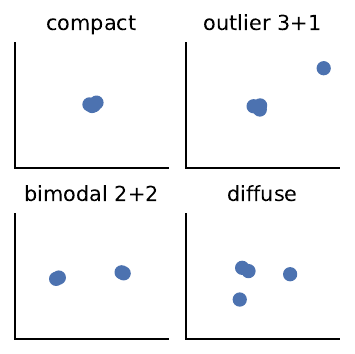}
\caption{Four configurations with similar mean spread but different structure (compact, $3{+}1$,
$2{+}2$, diffuse); a scalar maps them to one value.}
\label{fig:scalar}
\end{subfigure}\hfill
\begin{subfigure}[t]{0.58\textwidth}
\includegraphics[width=\textwidth]{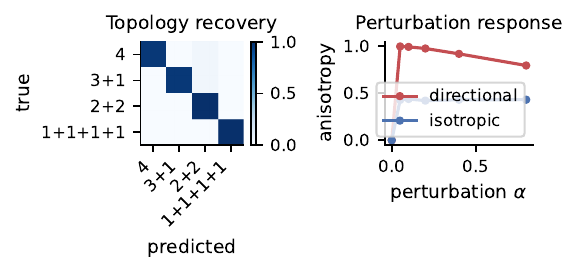}
\caption{Controlled recovery (Exp.~2): topology confusion matrix (left) and anisotropy response to
directional vs.\ isotropic perturbations (right).}
\label{fig:recovery}
\end{subfigure}
\caption{The profile separates seed set structures that a scalar cannot (a) and recovers them under
control (b).}
\label{fig:structure}
\end{figure}

On controlled seed sets of known geometry (compact, $3{+}1$, $2{+}2$, diffuse), the profile's
topology selector attains balanced accuracy \dTopoBalAcc{} and macro F1 \dTopoMacroF{} (chance
\dTopoChance{}), and its outlier gap component detects the $3{+}1$ configuration at AUROC
\dOutProfile{} where DCU reaches only \dOutDcu{} (Table~\ref{tab:recovery},
Figure~\ref{fig:recovery}). Under graded perturbations, anisotropy rises with directional noise
($\rho=\dAnisDir{}$) but far less with isotropic noise ($\rho=\dAnisIso{}$), exactly as intended. The
lone weak point is the symmetric two mode case: the mode ratio component detects $2{+}2$ at AUROC
\dMMProfile{} only (DCU \dMMDcu{}), which we report rather than conceal. \emph{Conclusion:} the
profile recovers outlier and directional structure well and topology as a whole very well, while
symmetric bimodality at $K{=}4$ remains hard.

\begin{table}[t]
\centering\small
\begin{minipage}[t]{0.5\textwidth}\centering
\caption{Controlled structural recovery (Exp.~2): the profile classifies topology and detects the
outlier configuration far above chance; a dispersion scalar (DCU) cannot.}
\label{tab:recovery}
\begin{tabular}{lrr}
\toprule
Metric & Value & Chance \\
\midrule
Topology balanced accuracy & 0.98 & 0.25 \\
Topology macro F1 & 0.98 & 0.25 \\
Outlier AUROC (outlier gap) & 1.00 & 0.50 \\
Outlier AUROC (DCU) & 0.35 & 0.50 \\
Multimodality AUROC (mode ratio) & 0.49 & 0.50 \\
Multimodality AUROC (DCU) & 0.64 & 0.50 \\
\bottomrule
\end{tabular}

\end{minipage}\hfill
\begin{minipage}[t]{0.46\textwidth}\centering
\caption{Incremental value (Exp.~7): CV $R^2$ for cross seed semantic disagreement. Distributional
structure (M2 to M6) does not beat plain pairwise dispersion (M0).}
\label{tab:incremental}
\begin{tabular}{lr}
\toprule
Model & CV $R^2$ \\
\midrule
M0 pairwise & 0.243 \\
M1 scalar uq & 0.289 \\
M2 simple geometry & 0.237 \\
M3 dist structure & 0.235 \\
M4 full sourceblind & 0.294 \\
M5 ref simple & 0.256 \\
M6 full profile & 0.240 \\
\bottomrule
\end{tabular}

\end{minipage}
\end{table}

\subsection{RQ3: What does the uncertainty track?}
\begin{figure}[t]
\centering
\includegraphics[width=\textwidth]{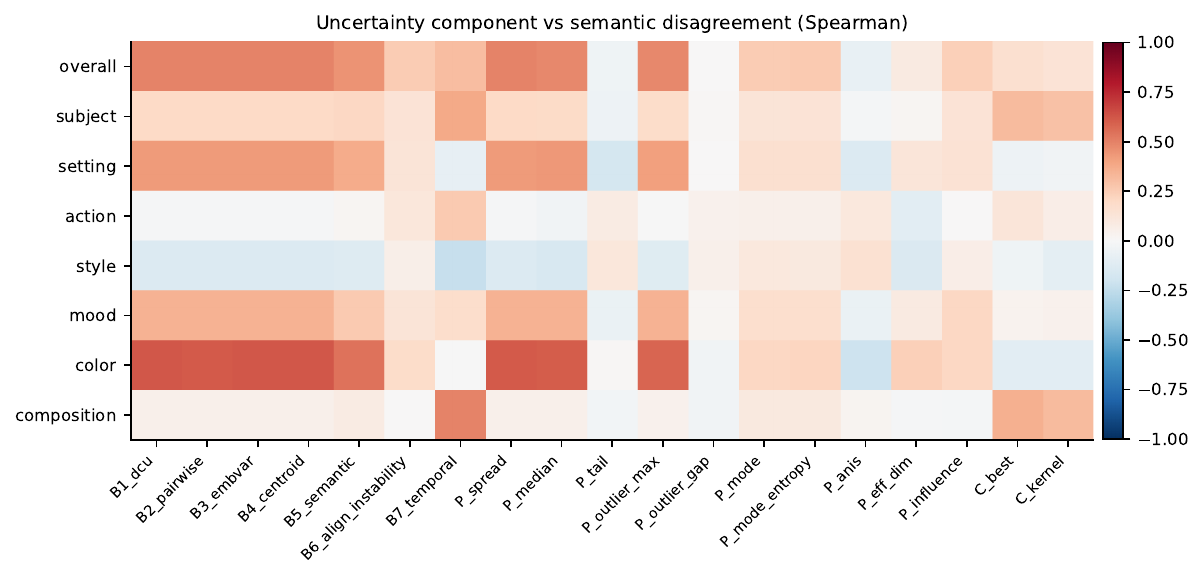}
\caption{Spearman correlation of each uncertainty component (columns) with per attribute cross seed
semantic disagreement (rows).}
\label{fig:semantic}
\end{figure}

Against structured cross seed semantic disagreement, the dispersion estimators are the strongest
single correlates (DCU $\rho=\dSemDcu{}$), and pairwise dispersion identifies the most divergent
quartile at AUROC \dSemClsPair{} (Figure~\ref{fig:semantic}). The purely structural components
(outlier gap, mode ratio, anisotropy) carry little \emph{additional} linear signal for this target.
\emph{Conclusion:} source blind uncertainty tracks observable semantic disagreement, and does so
chiefly through its dispersion magnitude.

\subsection{RQ4 and RQ5: Reference coverage and its separation from uncertainty}
The best seed reference similarity exceeds the mean by \dGapPoint{} (95\% CI
$[\dGapLo{},\dGapHi{}]$), so some seed sets already contain a faithful rendering that the mean
obscures. Kernel inclusion, however, is nearly identical to mean similarity ($\rho=\dKernMean{}$) and
to best of $K$ ($\rho=\dKernBest{}$) on this corpus, adding little beyond them; only the parametric
vMF likelihood departs from it ($\rho=\dVmfKern{}$). Seed uncertainty and reference coverage form
separate axes: the two axis taxonomy (Table~\ref{tab:taxonomy}) divides high uncertainty cases into
reference \emph{covering} diversity ($n{=}\dQcovDiv{}$) and reference \emph{missing} diversity
($n{=}\dQmissDiv{}$). \emph{Conclusion:} high diversity is not a single phenomenon, and the protocol
distinguishes meaningful variation from a failure to cover the target.

\begin{table}[t]
\centering\small
\caption{Two axis taxonomy (Exp.~6): high seed uncertainty splits into reference covering and
reference missing diversity.}
\label{tab:taxonomy}
\begin{tabular}{lrrrr}
\toprule
Quadrant & $n$ & Sem.\ disagr. & Mean sim. & Best sim. \\
\midrule
Low unc., high cov. & 60 & 0.57 & 0.45 & 0.50 \\
High unc., high cov. & 97 & 0.65 & 0.45 & 0.51 \\
Low unc., low cov. & 37 & 0.53 & 0.34 & 0.39 \\
High unc., low cov. & 56 & 0.64 & 0.32 & 0.37 \\
\bottomrule
\end{tabular}

\end{table}

\subsection{RQ6: Does the profile beat a single scalar?}
This is the strongest negative test. Nested models from pairwise only (M0) to the full distributional
profile (M6) predict cross seed semantic disagreement at cross validated $R^2$ of \dSemRtwoBase{}
(M0) and \dSemRtwoFull{} (M6): the full profile does \emph{not} improve on the single scalar
($\Delta R^2=\dSemDelta{}$), and the primary M6 against M5 comparison is likewise flat
($\Delta=\dMsixDelta{}$, 95\% CI $[\dMsixLo{},\dMsixHi{}]$; Table~\ref{tab:incremental}).
\emph{Conclusion:} for \emph{predicting} semantic disagreement, distributional structure is
diagnostic rather than additive, in keeping with the redundancy established in RQ1.

\subsection{RQ7 and RQ8: Reliability at four seeds and across encoders}
The simple estimators are reliable from subsets: three of four seed rankings reproduce the full four
ranking at Spearman \dRelThreePair{} (pairwise), \dRelThreeDcu{} (DCU), and \dRelThreeKern{} (kernel
coverage), and even two of four holds at \dRelTwoPair{} (Table~\ref{tab:seedcount}). Across the four
encoders (Table~\ref{tab:backbone}) the dispersion estimators agree at mean Spearman \dCBPair{}
(pairwise) and \dCBDcu{} (DCU), kernel coverage is moderately stable (\dCBKern{}), and the mode ratio
component is fragile (\dCBMode{}). \emph{Conclusion:} the simple components are trustworthy at
$K{=}4$ and robust across encoders, while the richer structural components demand more seeds.

\begin{table}[t]
\centering\small
\begin{minipage}[t]{0.5\textwidth}\centering
\caption{Seed count reliability (Exp.~8): Spearman agreement of subset estimates with the full four
seed ranking.}
\label{tab:seedcount}
\begin{tabular}{lrr}
\toprule
Estimator & 2 of 4 $\rho$ & 3 of 4 $\rho$ \\
\midrule
DCU & 1.00 & 1.00 \\
Pairwise & 1.00 & 1.00 \\
Emb.\ variance & 1.00 & 1.00 \\
Centroid & 1.00 & 1.00 \\
Semantic entropy & 0.97 & 0.98 \\
Best of $K$ & 0.98 & 0.99 \\
Kernel incl. & 1.00 & 1.00 \\
\bottomrule
\end{tabular}

\end{minipage}\hfill
\begin{minipage}[t]{0.46\textwidth}\centering
\caption{Cross backbone rank agreement over the four encoders (Exp.~9): mean and minimum Spearman
over encoder pairs.}
\label{tab:backbone}
\begin{tabular}{lrr}
\toprule
Estimator & Mean $\rho$ & Min $\rho$ \\
\midrule
Pairwise & 0.80 & 0.70 \\
DCU & 0.80 & 0.70 \\
Temporal & 0.94 & 0.92 \\
Kernel incl. & 0.70 & 0.57 \\
Outlier gap & 0.23 & 0.17 \\
Mode ratio & 0.35 & 0.22 \\
\bottomrule
\end{tabular}

\end{minipage}
\end{table}

\subsection{Temporal decomposition and a downstream use}
A temporal decomposition of global uncertainty into within video instability
($\overline{\phantom{x}}{=}\dTempWithin{}$), across seed same timestep disagreement (\dTempAcross{}),
and trajectory disagreement (\dTempTraj{}) shows that the three contribute unequally, with the
dominant source of global seed spread on this corpus being \emph{\dTempDom{}}. Finally, as one
concrete use, ranking artworks by predicted risk yields a selective prediction curve: on the
reference risk target the profile model improves over random by \dSelRefProfGain{} and
\dSelRefNBeat{} scorers beat random, while on the semantic risk target \dSelSemNBeat{} scorers beat
random, evidence that the identified structure is actionable rather than a policy contribution.

\section{Discussion}
The protocol accomplishes two things a scalar cannot. It \emph{names} the structure of a seed set
(compact, outlier, bimodal, diffuse, directional) and recovers that structure reliably (RQ2), and it
\emph{separates} seed uncertainty from reference coverage, resolving high diversity into meaningful
variation versus a failure to cover the target (RQ4, RQ5). These are the diagnostic uses we
recommend.

Equally, the protocol is candid about where structure does \emph{not} help. Adding outlier, mode, and
anisotropy components does not improve the prediction of semantic disagreement over a single
dispersion scalar (RQ6), because on this corpus the source blind estimators span largely one axis
(RQ1), and source blind uncertainty is interpretive rather than reconstructive, measuring how the
seeds vary rather than their fidelity to the withheld artwork. A practitioner should therefore deploy
the profile to \emph{classify} uncertainty and route decisions, but should not expect a richer scalar
to \emph{predict} meaning better than pairwise dispersion, nor any source blind score to substitute
for reference fidelity.

\section{Conclusion}
We presented the first systematic study of the structure of generative uncertainty for modern art
animation, with a protocol (baseline suite, distributional profile, distribution model ablation,
eight identification questions, and an artwork level statistical protocol) applied end to end to a
purpose built corpus of \dNVid{} source blind videos across \dNEnc{} encoders. As a diagnostic it
recovers seed set topology, isolates outliers, and separates reference covering from reference
missing diversity, reliably at four seeds and across encoders, while two negative results mark its
limits: structure does not beat a single scalar for semantic disagreement, and source blind
uncertainty is interpretive rather than reconstructive. Both the protocol and the corpus are
released.

\footnotesize
\bibliographystyle{plainnat}
\bibliography{references}

\end{document}